\documentclass[aps, reprint, superscriptaddress]{revtex4-2}
\usepackage[utf8]{inputenc}
\usepackage{graphicx}
\usepackage{amsmath, amssymb}
\usepackage{xcolor}
\usepackage{upgreek}
\usepackage{bm}
\usepackage{hyperref}
\usepackage{siunitx}

\newcommand{\D}{\mathrm{d}}
\newcommand{\ssb}{\mathbf{s}}
\newcommand{\Eb}{\mathbf{E}}
\newcommand{\Bb}{\mathbf{B}}
\newcommand{\vb}{\mathbf{v}}
\newcommand{\Omegab}{\bm{\Omega}}

\begin{document}

	\title{A plasma photocathode for spin-polarized electron beams via state-selected hydrogen halide photofragments}
	
	\author{Lars Reichwein}
	\email{l.reichwein@fz-juelich.de}
    \affiliation{Peter Gr\"{u}nberg Institut (PGI-6), Forschungszentrum J\"{u}lich, 52425 J\"{u}lich, Germany}
	\affiliation{Institut f\"{u}r Theoretische Physik I, Heinrich-Heine-Universit\"{a}t D\"{u}sseldorf, 40225 D\"{u}sseldorf, Germany}

    \author{Thomas C. Wilson}
    \affiliation{Institut f\"{u}r Theoretische Physik I, Heinrich-Heine-Universit\"{a}t D\"{u}sseldorf, 40225 D\"{u}sseldorf, Germany}

    \author{Dimitris Sofikitis}
    \affiliation{Department of Physics, Atomic and Molecular Physics Laboratory, University of Ioannina, University Campus, 45110 Ioannina, Greece}

    \author{Chinmaya Singh}
    \affiliation{Department of Physics, University of Crete, 70013 Heraklion-Crete, Greece}
    \affiliation{Institute of Electronic Structure and Lasers, Foundation for Research and Technology-Hellas, 71110 Heraklion-Crete, Greece}

    \author{T. Peter Rakitzis}
    \affiliation{Department of Physics, University of Crete, 70013 Heraklion-Crete, Greece}
    \affiliation{Institute of Electronic Structure and Lasers, Foundation for Research and Technology-Hellas, 71110 Heraklion-Crete, Greece}

    \author{Bernhard Hidding}
	\affiliation{Institut f\"{u}r Laser- und Plasmaphysik, Heinrich-Heine-Universit\"{a}t D\"{u}sseldorf, 40225 D\"{u}sseldorf, Germany}

	\author{Alexander Pukhov}
	\affiliation{Institut f\"{u}r Theoretische Physik I, Heinrich-Heine-Universit\"{a}t D\"{u}sseldorf, 40225 D\"{u}sseldorf, Germany}

    \author{Liangliang Ji}
    \affiliation{State Key Laboratory of Ultra-intense Laser Science and Technology, Shanghai Institute of Optics and Fine Mechanics, Chinese Academy of Sciences, Shanghai 201800, People’s Republic of China}

    \author{Markus B\"{u}scher}
	\affiliation{Peter Gr\"{u}nberg Institut (PGI-6), Forschungszentrum J\"{u}lich, 52425 J\"{u}lich, Germany}
	\affiliation{Institut f\"{u}r Laser- und Plasmaphysik, Heinrich-Heine-Universit\"{a}t D\"{u}sseldorf, 40225 D\"{u}sseldorf, Germany}
    
	\date{\today}
	
	\begin{abstract}
    Spin-polarized electron beams are essential tools for probing fundamental symmetries and for the search beyond the Standard Model. While plasma-based accelerators are a promising pathway towards higher-energy frontiers, they have so far failed to deliver a competitive polarized source: existing proposals are challenging to realize and achievable polarizations remain far below conventional sources. Here, we introduce a photocathode-like scheme, applied to a gas of pre-polarized hydrogen and halogen atoms. A VUV and a visible laser pulse excite the halogen atoms to create a two-component ionization medium, consisting of low-threshold excited halogen atoms and high-threshold polarized hydrogen. Particle-in-cell simulations show witness beams with tens of pC charge retaining up to 97\% of the initial polarization, rivaling state-of-the-art conventional sources.
	\end{abstract}
	
	\maketitle


    Various applications in high-energy physics require spin-polarized beams \cite{Glashausser1979, Chen2026}, typically $P \geq 70-80\%$.  In conventional rf-based accelerators, such polarized electron beams are commonly generated using photoguns \cite{Litvinenko2026}.
    While these are readily available in dedicated accelerator laboratories, the wakefield community only has very limited access to such devices. The long-term goal of that community is nevertheless a wakefield-based collider at the 10 TeV pCM frontier \cite{Gessner2025}. 
    In the context of wakefield acceleration, polarized beams have attracted attention only more recently. Previous studies mainly focused on optimizing parameters like beam energy \cite{Picksley2024} or beam charge \cite{Goetzfried2020}. A key development towards ultra-low emittance beams has been the proposal of the plasma photocathode (also dubbed ``trojan horse regime''), in which a two-component gas with well-separated ionization thresholds allows controlled injection into the wake \cite{Hidding2012, Habib2023}.

    An overview of the state-of-the-art for plasma-based, polarized beams was recently given in Ref. \cite{Reichwein2025}.
    The only proof-of-principle experiment on polarized beams from laser-plasma interaction thus far has been on nuclear-spin-polarized helium-3 \cite{Zheng2026}.
    For polarized electrons, two routes are being studied theoretically: (i) initially unpolarized targets, where a polarized witness beam is created in-situ and (ii) the pre-polarized case, where the target needs to be prepared before the actual wakefield acceleration and polarization maintained during injection.
    
    For the former case, Nie \textit{et al.} have proposed a plasma photocathode setup relying on the ionization of the ytterbium $4f^{14}$ orbital using circularly polarized (CP) laser pulses \cite{Nie2022}. The scheme is experimentally demanding, and the simulated 4 kA beams reach only $\sim 56\%$ polarization, which is below the requirements for future high-energy physics applications.

    The latter, pre-polarized case relies on the photo dissociation of hydrogen halides \cite{Sofikitis2025}. Here, the initial degree of polarization can be much higher, but the attainable target density and interaction volume have remained challenging for wakefield acceleration schemes thus far. One crucial problem has been to avoid the injection of unpolarized electrons from the halogen component which necessitated the formation of a channel containing only polarized hydrogen. Aligning the wakefield with this small channel would be challenging to realize. In Ref. \cite{Reichwein2026}, pinching of a mismatched electron driver was proposed as a means of localized injection. This relaxes some target constraints but still yields only $P \sim 50\%$.
    
    In this paper, we propose a modified setup based on hydrogen halides which removes the limitation of the narrow plasma channel and provides high-quality electron beams. Using additional VUV and visible laser pulses to excite the halogen component, its first ionization energy is significantly reduced, enabling a plasma photocathode to be driven directly in the hydrogen halide mixture. Our particle-in-cell (PIC) simulations show the polarization to exceed $90\%$ for witness beams of tens of pC.

    \begin{figure*}[th]
        \centering
        \includegraphics[width=\textwidth]{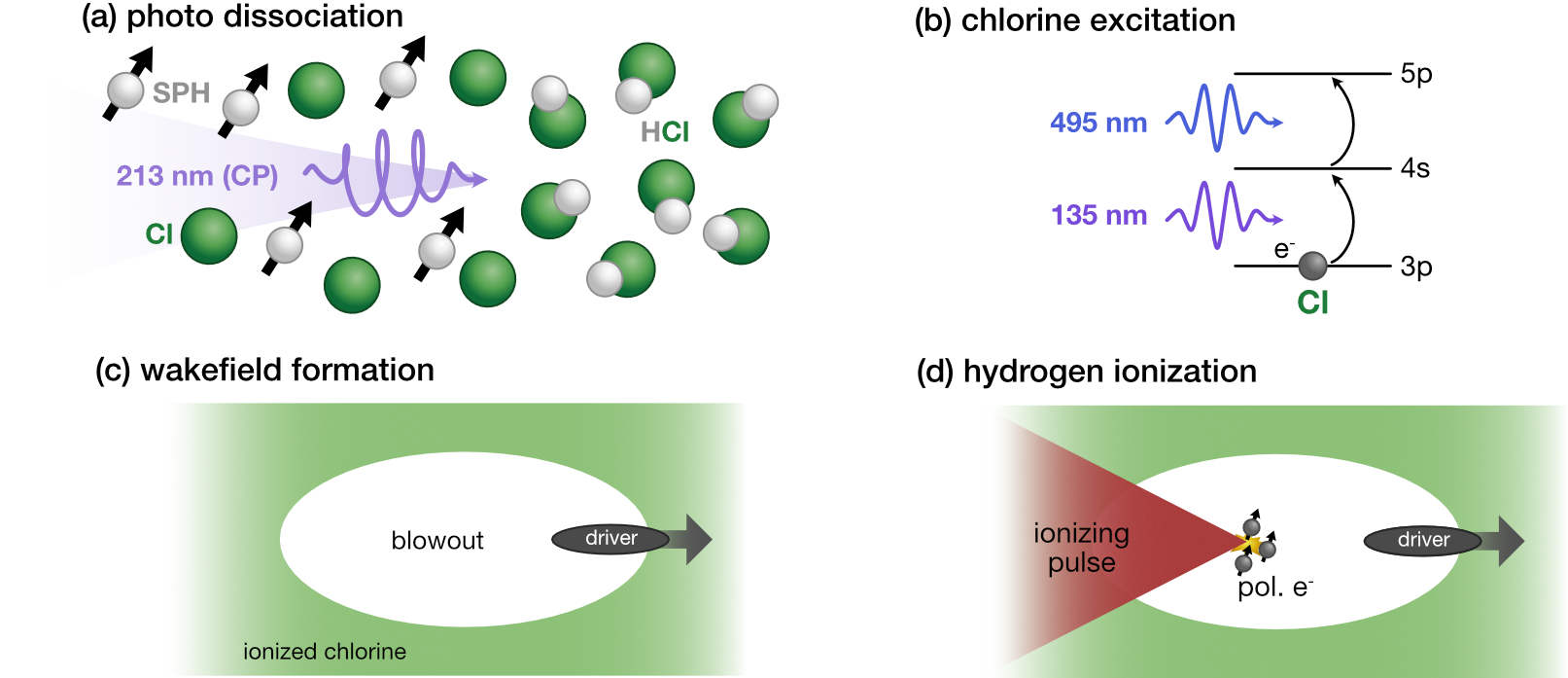}
        \caption{\label{fig:scheme}Schematic of the proposed scheme: (a) The first UV laser dissociates the HCl molecules and imparts polarization on the hydrogen component. (b) The second UV laser and visible laser excite the chlorine electrons. (c) A relativistic electron beam generates a wakefield only using the chlorine electrons ($\mathrm{Cl}^{1+}$). (d) A weak laser pulse locally ionizes the hydrogen, releasing polarized electrons inside the wakefield.}
    \end{figure*}
    
    \section*{Results}
    \subsection*{Concept and physical mechanism of polarized electron generation}
    Previous halide-based schemes to produce spin-polarized electrons for wakefield acceleration required the formation of a channel devoid of halogen atoms, containing only the polarized protons/electrons \cite{Sofikitis2025}. The achievable channel density and volume strongly constrain state-of-the-art wakefield injection schemes and require fine alignment of the wake with the channel position.

    In the present scheme (cf. Fig. \ref{fig:scheme} for a schematic), no such channel is necessary, and an essentially homogeneous density profile can be used.
    Starting point of the scheme remains the photo dissociation of the hydrogen halide (e.g. HCl) with a circularly polarized UV (213 nm) laser pulse. Upon dissociation the hydrogen component gains its polarization, and the fragments separate. Given sufficient time, the separating fragments form a channel in the target (which was utilized in the setup of \cite{Sofikitis2025}). Instead of waiting for the channel to fully form, we subsequently employ a separate VUV laser at 135 nm and a visible laser pulse at around 497 nm to excite the chlorine atoms to the $5p$ state, via the first electronic excited state, i.e. the $3p \to 4s$ transition. This excitation using ``stimulated Raman adiabatic passage'' (STIRAP) can transfer 100\% of the chlorine population to the excited state \cite{Bergmann1998, Vitanov2017}. The lifetime of the final excited state is on the order of 1 {\textmu}s, for the intermediate states it is 10-20 ns.

    The excitation effectively lowers the first ionization level of chlorine from 12.97 eV to approx. 1.21 eV, thus making the gap to the hydrogen ionization energy (13.6 eV) large enough to drive a plasma photocathode.
    Using a high-energy electron beam, a wakefield can be driven solely in $\mathrm{Cl}^{1+}$, utilizing the space-charge fields of the beam to ionize the gas. This forms the low ionization-threshold (LIT) component of the scheme. The polarized hydrogen component remains non-ionized at this stage.
    For the presented scheme, the exact energy of ionization is not relevant as long as the ionization levels of chlorine and hydrogen are sufficiently separated; the generated wake would only differ marginally.
    Using different laser wavelengths, the chlorine could alternatively be excited to other levels, e.g. 2.72 eV for the first ionization energy.

    A subsequent, weak ($a_0 \ll 1$) laser pulse with 800 nm is used to locally ionize electrons from the spin-polarized hydrogen component (the high-ionization threshold [HIT] component) directly inside the wakefield. 
    As with the conventional, unpolarized plasma photocathode, these electrons fall back within the blowout while gaining longitudinal momentum, before being trapped by the wakefield.

     The exact degree of initial polarization still depends on the target density and volume, as well as the selected hydrogen halide (cf. Ref. \cite{Sofikitis2025}). For example, choosing $n_\mathrm{Cl} = 10^{17}$ cm$^{-3}$, the attainable target diameter becomes approx. 100 {\textmu}m for an initial polarization of $ \lesssim 70\%$. The initial degree of polarization can be vastly improved using an additional infrared laser for bond alignment, bringing it to $P\sim 100\%$ \cite{Spiliotis2021}. The excitation step can similarly be realized with other halogens (see Methods section).
     
    \subsection*{PIC demonstration}

    To demonstrate the scheme, we conduct particle-in-cell simulations using the code \textsc{fbpic} \cite{fbpic} which contains a description of spin precession according to the T-BMT equation \cite{Thomas1926, Bargmann1959}. 
    We model the electron beam driver with energy $\SI{200}{\mega\electronvolt}\pm 10\%$, containing \SI{480}{\pico\coulomb} total charge. The driver is a Gaussian ellipsoid with dimensions $\sigma_z = \SI{7}{\micro\metre}$ and $\sigma_r = \SI{4}{\micro\metre}$, and normalized emittance $\epsilon_n =\SI{20}{\micro\metre}$. Such a drive beam could be produced via laser-driven wakefield acceleration (cf. for example Ref. \cite{Couperus2017}).
    The plasma is a uniform slab consisting of a mixture of unpolarized, excited chlorine atoms at $n_{\mathrm{Cl}}=\SI{1e17}{\per\cubic\centi\metre}$, and polarized hydrogen atoms at $n_{\mathrm{H}}=\SI{1e16}{\per\cubic\centi\metre}$. This density ratio is chosen to account for the fact that after some fly-out time, the hydrogen density in the interaction volume will be lower than that of chlorine. The effect of varying this ratio is studied in the later sections. Initially, both species are completely non-ionized.
    The electrons associated with the spin-polarized hydrogen (SPH) component are initially polarized entirely along the longitudinal axis ($s_z \equiv 1$). The plasma begins at $z=0$ and has a linear up-ramp of length \SI{335}{\micro\metre}.

    The ionizing laser is modeled as a linearly polarized Gaussian pulse with wavelength $\lambda_0 = \SI{800}{\nano\metre}$, amplitude $a_0 = 0.01$, duration $\tau = \SI{33}{\femto\second}$ and spot size $w_0 = \SI{8}{\micro\metre}$. The laser is focused to approximately \SI{150}{\micro\metre} after the end of the up-ramp ($z_f = \SI{500}{\micro\metre}$) and trails behind the driver with a delay of \SI{75}{\femto\second} (\SI{22}{\micro\meter} on the grid).

    \begin{figure}[h]
        \includegraphics[width=\columnwidth]{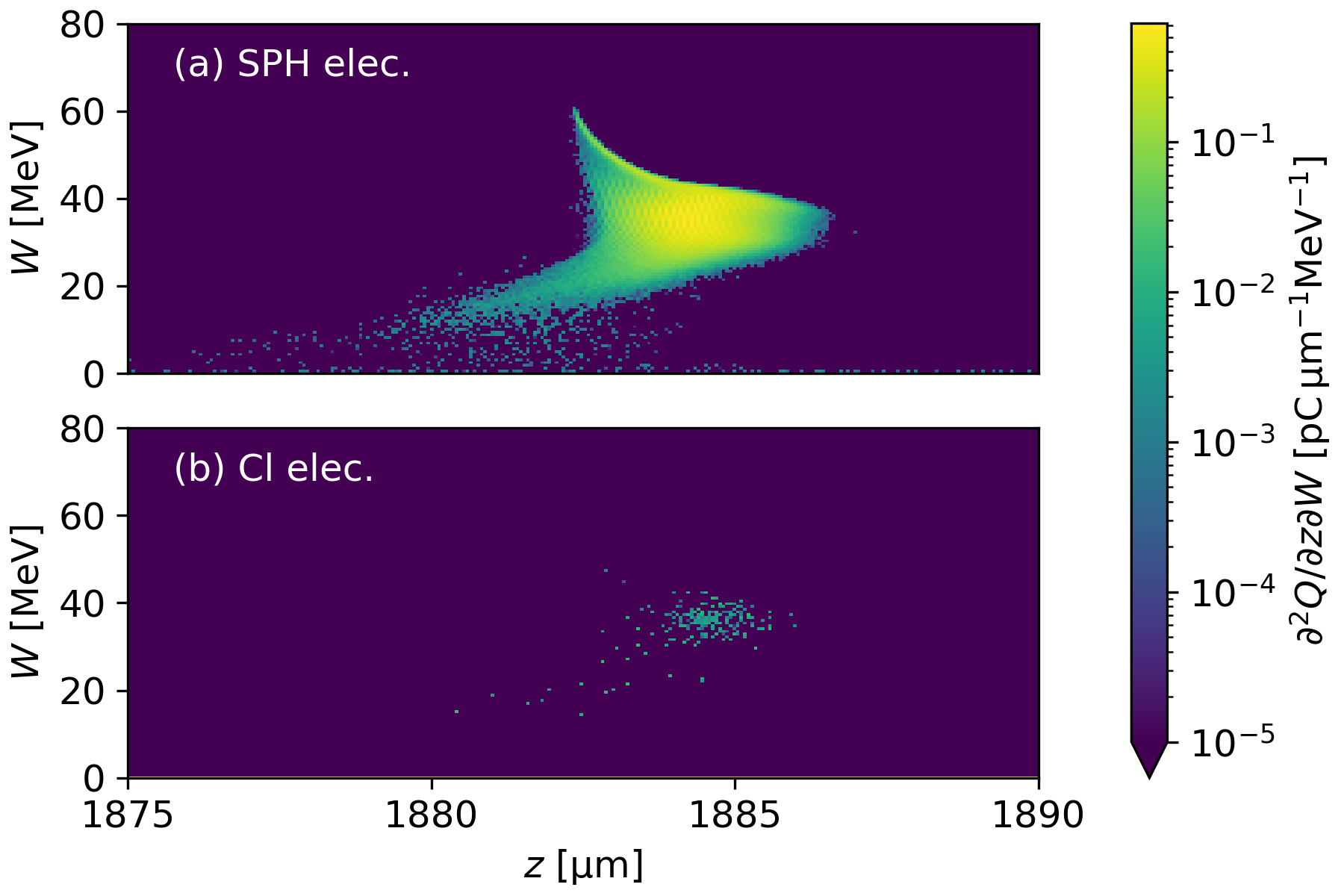}
        \centering
        \caption{\label{fig:fbpic_phase} Phase-space comparison for (a) the SPH electrons and (b) the chlorine electrons. Only a small amount of unpolarized chlorine electrons is present in the same phase-space volume, thus affecting total beam polarization only marginally.}
    \end{figure}

\begin{figure}[h]
        \includegraphics[width=\columnwidth]{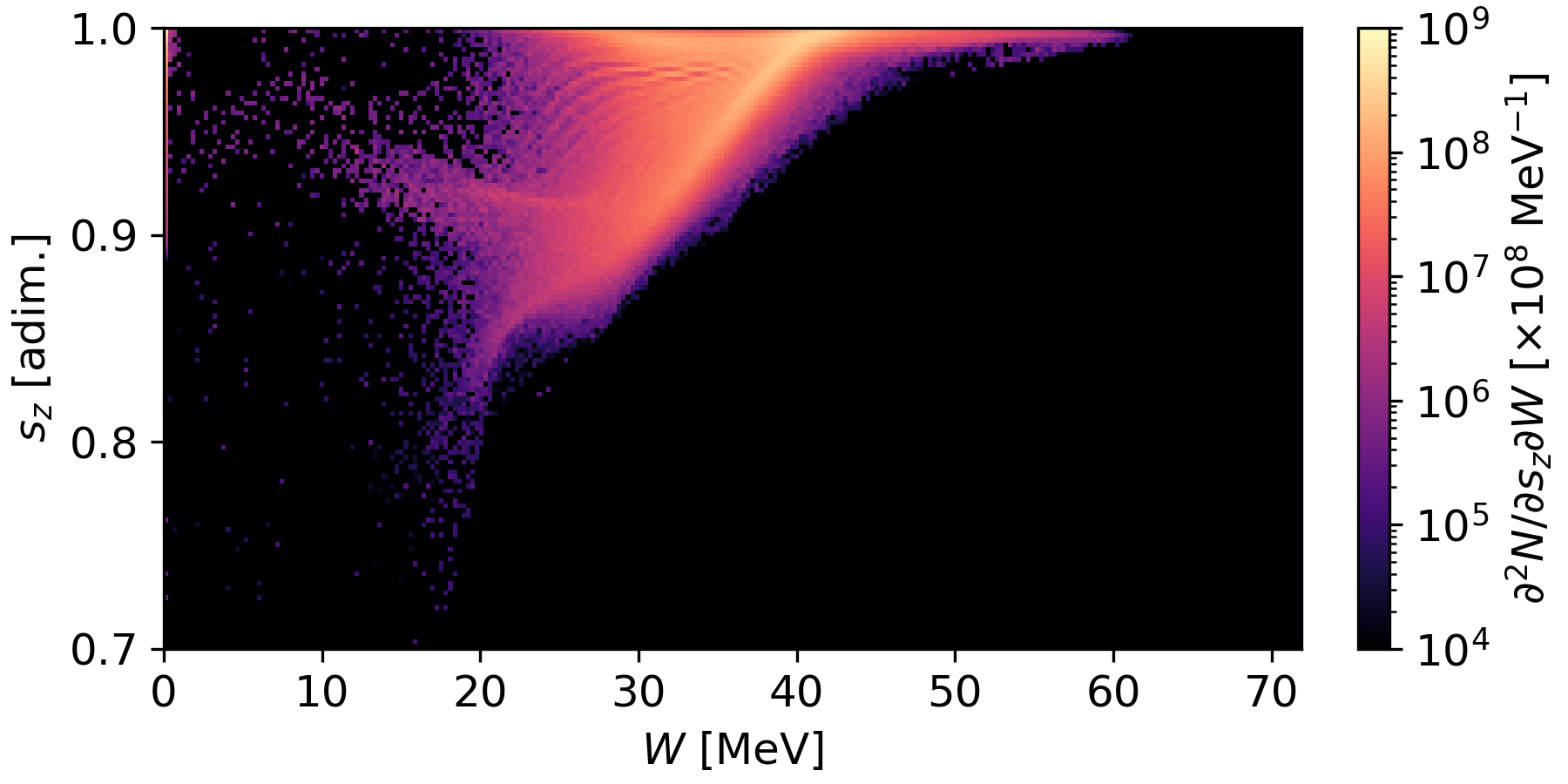}
        \centering
        \caption{\label{fig:fbpic_pol} Polarization phase-space of the SPH electrons after 1.3 mm of acceleration. Note that the highest-energy electrons exhibit the highest polarization.}
    \end{figure}

    The simulation results in Fig. \ref{fig:fbpic_phase} show the final phase spaces for electrons from both the SPH and Cl species after 1.3 mm of propagation, with the SPH contribution dominant. Excluding particles with energies less than \SI{5}{\mega\electronvolt} and those beyond a radius of \SI{10}{\micro\metre}, the witness contains \SI{10.8}{\pico\coulomb} charge, with energy $\SI{35.2}{\mega\electronvolt}\pm 14.8\%$ and average normalized emittance $\sqrt{\epsilon_x\epsilon_y} = \SI{57}{\nano\metre}$. The final polarization phase-space of the witness is shown in Fig. \ref{fig:fbpic_pol}, and shows that the highest-energy particles have the highest polarization degree. The witness beam has an average polarization of 97\% (of the initial value). The essentially unpolarized chlorine electrons contribute only \SI{0.04}{\pico\coulomb} to the witness bunch, which represents a negligible degradation of the overall polarization.

    \subsection*{Robustness and parameter dependence}

   The 97\% polarization quoted above assumes a fully polarized initial population of hydrogen, $P_0 \sim 100\%$, before injection. As extensively discussed in Ref. \cite{Sofikitis2025}, depending on the target density and interaction volume the initial polarization value can be significantly lower than that.
   In the proposed density regime, where $n_\mathrm{Cl} = 10^{17}$  cm$^{-3}$ and $n_\mathrm{H} = 10^{16}$  cm$^{-3}$, the previously described driver generates a wakefield of approx. 80 {\textmu}m diameter. Given this volume requirement, an HCl dissociation target would yield 70\% initial polarization, such that our scheme offers $ P = 0.97 P_0 = 67.9\%$. This value is already close to the necessary 70\% for HEP applications. It can, however, be substantially improved by aligning the bonds of the HCl molecules: employing a separate infrared laser pulse can improve the initial polarization to essentially 100\% \cite{Spiliotis2021}. In this case, our scheme would indeed produce 97\% final polarization.
   While adding an alignment laser makes the scheme more complex, the improvements in polarization and simplified target geometry make up for it: the wakefield acceleration does not need to be carefully aligned with a narrow channel as in Ref. \cite{Sofikitis2025} and the polarization yield is competitive with state-of-the-art conventional sources.

\begin{figure}[h]
        \centering
        \includegraphics[width=\columnwidth]{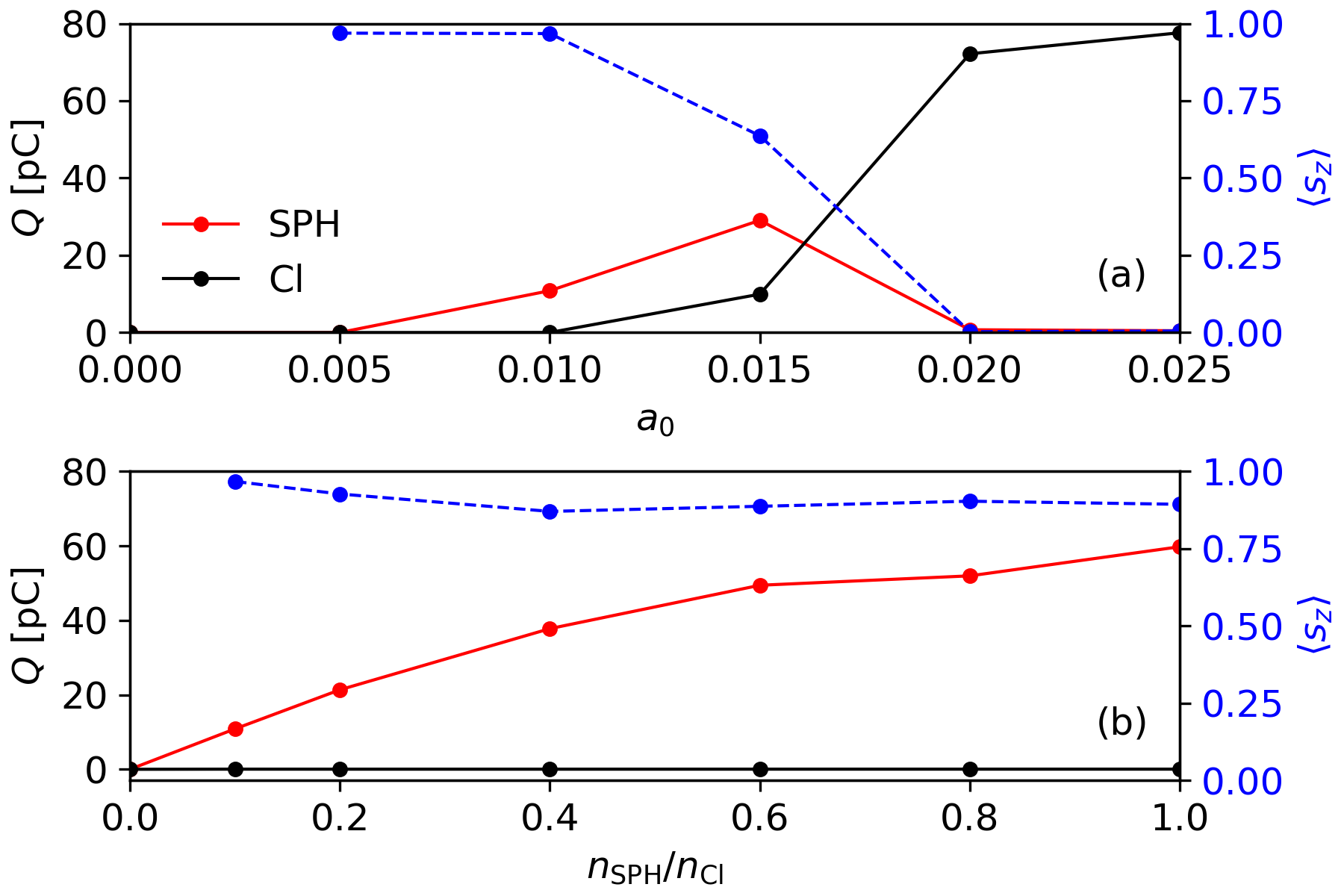}
        \caption{\label{fig:param_scan} Dependency of the accelerated charge and average polarization against (a) ionizing laser amplitude $a_0$, and (b) density ratio $n_\mathrm{SPH} / n_\mathrm{Cl}$. The red and black points denote the SPH and chlorine electrons, respectively, with the average polarization of the entire witness (Cl + SPH) shown in blue (right hand scale).}
    \end{figure}

    The attainable witness beam charge can be tuned by employing different laser and/or target parameters. In Fig. \ref{fig:param_scan} we show the dependence of beam charge and polarization on the ionizing laser amplitude $a_0$ as well as the hydrogen-to-chlorine density ratio.
    
    The choice of $a_0$ is effectively limited by the fact that ionization to Cl$^{2+}$ ($\sim 23.81$ eV) needs to be avoided in order not to introduce unwanted, unpolarized electrons. We observe in Fig. \ref{fig:param_scan}(a) that for $a_0 > 0.01$, chlorine ($\mathrm{Cl}^{2+}$) electrons begin to be ionized and trapped in appreciable amounts, causing the average polarization to drop significantly. For high $a_0 > 0.02$ the volume of trapped chlorine electrons is now so large that it interferes with the trapping of the SPH electrons, and the witness becomes completely unpolarized.

    Since the delay between dissociation and the subsequent steps is a free parameter, the density ratio can be tuned continuously from $n_\mathrm{H} = n_\mathrm{Cl}$ (directly after dissociation) to $ n_\mathrm{H} \ll n_\mathrm{Cl}$. Figure \ref{fig:param_scan}(b) shows the effect of varying the ratio of SPH to background Cl atoms for fixed $a_0 = 0.01$. We see that there is a simple dependence, where a larger ratio translates to higher trapped charge ($\sim 60$ pC with 90\% polarization). The amount of trapped chlorine electrons remains negligible at all times, and the overall bunch polarization remains high. The trade-off for high SPH density is that the bunch quality suffers, with the witness emittance rising from $\sim\SI{50}{\nano\metre}$ when witness charge is very low, to $\sim\SI{1.5}{\micro\metre}$ when charge is high.

    The electron driver's space-charge field itself is another factor to be considered: the driver parameters need to be chosen such that it, ideally, ionizes only chlorine electrons while maintaining a strong enough wake to trap SPH electrons. While a higher-charge driver would technically drive a stronger wake, it would also ionize parts of the SPH component. These additional electrons will generally not be trapped by the wake, but the volume of neutral SPH available for trapping is reduced accordingly.
    
    Independently of spin, the ionization should be aligned with $r = 0$ of the blowout to minimize the final emittance. In the presented case, having transverse offsets will also induce stronger spin precession since the precession frequency $|\Omegab|$ scales linearly with $\Eb, \Bb$ and the transverse electromagnetic fields inside the blowout scale linearly with the radial offset.

    Finally, the eventual applications of such polarized beams for particle physics will require significantly higher energies than the $\sim 35$ MeV over 1.3 mm we have discussed for our scheme here.
    
    The continued acceleration of these beams need not rely on a hydrogen halide target. Instead, a secondary stage of wakefield acceleration with higher gradients and entirely different target composition may be used. 
    Since the prefactors of the precession frequency $\Omega_B, \Omega_E \to a \approx 10^{-3}$ for $\gamma \gg 1$, the beam polarization will be approximately preserved, as was also seen in the studies of Wu \textit{et al.} \cite{Wu2019pwfa, Wu2019lwfa} which considered a density down-ramp scheme for highly idealized HCl target parameters, as well as the in-situ scheme by Nie \textit{et al.} \cite{Nie2021, Nie2022}.

    Since the prefactors of the precession frequency $\Omega_B, \Omega_E \to a \approx 10^{-3}$ for $\gamma \gg 1$, the beam polarization will be approximately preserved, as was also seen in the studies of Wu \textit{et al.} \cite{Wu2019pwfa, Wu2019lwfa} which considered a density down-ramp scheme for highly idealized HCl target parameters, as well as the in-situ scheme by Nie \textit{et al.} \cite{Nie2021, Nie2022}. This behavior can also be observed in the analytical model of Vieira \textit{et al.} \cite{Vieira2011}.
  
    As discussed by Mathiak \textit{et al.} in Ref. \cite{Mathiak2026}, radiative effects should be of negligible contribution for staging wakefields towards high-energy applications. An extended discussion of quantum electrodynamical limitations for polarized beams towards high energies has also been given by Qian \textit{et al.} \cite{Qian2026}.

    \section*{Discussion}
    We have proposed a scheme to obtain high-polarization witness beams from pre-polarized hydrogen halides (e.g. HCl). The halogen component is excited using a VUV laser pulse and a visible laser pulse reducing the first ionization energy of chlorine to approx. 1.21 eV in the case of chlorine. This enables us to drive a plasma photocathode, where the excited halogen is used for wakefield formation and the pre-polarized hydrogen component for subsequent, localized ionization. The resulting witness beams exhibit tens of pC charge and a polarization exceeding 90\% of the initial value.

    Compared to previous schemes, the present one alleviates the restrictions of channel formation and the accompanying alignment challenges. Moreover, the polarization values are high enough to be of use for high-energy physics applications and rival conventional sources. One crucial advantage of such plasma-based, polarized sources is that they can be realized in laboratories equipped for laser-matter interaction which rarely have access to conventional photoguns.
    
    Beyond the immediate application to polarized sources, the concept of excited states for plasma photocathodes can be of interest for improved separation between the LIT and HIT components, and thus improved injection control and witness quality.
    Experimental verification of the proposed scheme is envisioned at the Shanghai Institute of Optics and Fine Mechanics for the near future.

    \section*{Methods}

    \subsection*{Particle-in-cell simulations}

    We conduct particle-in-cell simulations using the code \textsc{fbpic} \cite{fbpic} which employs cylindrical geometry with azimuthal mode decomposition, and features a pseudo-spectral field solver and GPU acceleration, allowing fast and accurate simulations of wakefield acceleration.
    The simulation is executed in axisymmetric mode, using only a single azimuthal mode ($N_m = 1$). We use a grid resolution of $\Delta z \approx \SI{31}{\nano\metre}$, $\Delta r \approx \SI{62}{\nano\metre}$. The simulation is evolved for 64{\,}000 iterations (\SI{6.6}{\pico\second}), giving the witness approximately \SI{1.3}{\milli\metre} of acceleration.

    The ionizing laser is modeled analytically and the fields are applied directly to the particle species. This approach is justified as the laser amplitude and plasma density are both low, $a_0 \ll 1$, $n_e \ll n_c$, and so nonlinear effects on the laser propagation will be negligible. \textsc{fbpic} uses the DC-ADK model of ionization \cite{Chen2013}.

    \subsection*{Modeling of spin dynamics}
    Spin evolution in \textsc{fbpic} is solely based on the T-BMT equation \cite{Thomas1926, Bargmann1959}: Upon ionization, the electron spin precesses according to
    \begin{align}
        \frac{\D \ssb}{\D t} = - \Omegab \times \ssb \; .
    \end{align}
    Here, 
    \begin{align}
        \Omegab = \frac{e}{mc} \left[ \Omega_B \Bb - \Omega_v \left( \frac{\vb}{c} \cdot \Bb \right) \frac{\vb}{c} - \Omega_E \frac{\vb}{c} \times \Eb \right] \; , \label{eq:prec}
    \end{align}
    is the precession frequency with $e$ denoting the charge, $m$ the mass, $c$ the vacuum speed of light, $\vb$ the particle velocity and $\Eb, \Bb$ the electromagnetic field. The prefactors are defined as
    \begin{align}
        \Omega_B = a + \frac{1}{\gamma} \; , && \Omega_v = \frac{a \gamma}{\gamma + 1} \; , && \Omega_E = a + \frac{1}{\gamma + 1} \; .
    \end{align}
    The factor $a = \alpha / (2\pi) \approx 10^{-3}$ is the electron's anomalous magnetic moment and $\gamma$ its Lorentz factor.

    Neither radiative effects nor Stern Gerlach-type forces are included, as their contributions in our parameter regime are negligible \cite{Thomas2020}. Given the parameters of the ionization pulse (in particular $a_0 \ll 1$), it is also reasonable to assume that the electron polarization will be preserved during the ionization process.

    \subsection*{Hydrogen-halide target configuration}

    When a circularly polarized UV laser pulse dissociates hydrogen halide molecules like HCl, the hydrogen component becomes polarized. Due to the hyperfine interaction, the polarization oscillates periodically between the proton and the electron with the hyperfine period $T \approx 0.7$ ns \cite{Sofikitis2025}, which is why these targets can also be used for the acceleration of polarized protons \cite{Reichwein2022}. The ionization and acceleration steps therefore need to be timed accordingly.
    Upon dissociation, the fragments will separate, leading to channel formation for long ``fly-out'' timescales. This separation has been utilized in Ref. \cite{Sofikitis2025} for the separation of the hydrogen carrying polarized electrons and the essentially unpolarized chlorine electrons. 
    
    While we have only described the use of HCl for our proposed scheme, it could be similarly conducted with other hydrogen halides. The use of HCl necessitates, as aforementioned, the use of a 135 nm VUV laser and a 495 nm visible laser for the specific excitation step. Excitation to other orbitals using the STIRAP approach would require different wavelengths. Assuming the ionization energy is still significantly below that of hydrogen, the photocathode steps would not be affected, since the driver is still capable of ionizing large volumes of the chlorine and thus drives the same wakefield.
    
    One alternative to HCl could, e.g.,  be the excitation of HI. As discussed in Ref. \cite{Sofikitis2025}, HI can similarly be dissociated for pre-polarization.  The excitation requires slightly different wavelengths depending on the orbital, a 183 nm laser for $2p_{3/2}$ and 206 nm for $2 p_{1/2}$. For the latter, approx. $10^{15}$ photons per pulse can be generated (compared to $\sim 10^{13}$ for the HCl case).
    Using different hydrogen halides changes the available density and volume range only marginally (discussed in \cite{Sofikitis2025}). 
    Without bond alignment, the initial polarization achievable by dissociation is limited by the required target density and volume. Using an additional IR laser pulse, the bonds can be aligned which brings the initial polarization towards $100\%$ for a broader parameter range \cite{Spiliotis2021}.
    
    \section*{Data availability}

    Source data required to reproduce the figures in this study are available from the authors upon reasonable request.

    \section*{Code availability}

    The particle-in-cell code \textsc{fbpic} is openly available \cite{fbpic}.

    \bibliography{excited_chlorine}

    \begin{acknowledgments}
         The authors gratefully acknowledge the Gauss Centre for Supercomputing e.V. \cite{GCS} for funding this project (spaf) by providing computing time through the John von Neumann Institute for Computing (NIC) on the GCS Supercomputer JUWELS at J\"ulich Supercomputing Centre (JSC). The work of M.B. has been carried out in the framework of the JuSPARC (J\"ulich Short-Pulse Particle and Radiation Center \cite{JuSPARC}).
         A.P. has been supported by BMBF (Project 05P24PF1).
         L.J. acknowledges the support by the National Science Foundation of China (Grant No. 12388102) and the Strategic Priority Research Program of Chinese Academy of Sciences (No. XDB0890302).
         T.P.R. and C.S. acknowledge support from the European Union’s Marie Sk{\l}odowska-Curie Actions under the EPACE programme (Grant Agreement No. 101169117).
    \end{acknowledgments}

    \section*{Author contributions}

    L.R. and T.P.R. conceived of the scheme. T.C.W. and L.R. conducted the PIC simulations with input from A.P.
    L.R. wrote the paper, with input from D.S., T.P.R., C.S., L.J., M.B. and B.H. All authors discussed the results and contributed to the final manuscript.

    \section*{Competing interests}

    The authors declare no competing interests.

\end{document}